\documentclass{cimento}

\begin{document}

\title{Spectroscopy of heavy baryons}

\author{Roelof Bijker\from{ins:x} \atque
Emmanuel Ortiz-Pacheco\from{ins:y}}
\instlist{\inst{ins:x} Instituto de Ciencias Nucleares, Universidad Nacional Aut\'onoma de M\'exico,\\
Coyocac\'an, 04510 Ciudad de M\'exico, Mexico
  \inst{ins:y} Jozef Stefan Institute, Jamova 39, 1000 Ljubljana, Slovenia}
  
\maketitle

\begin{abstract}
We present a study of mass spectra and electromagnetic couplings of $S$- and $P$-wave heavy baryons in the framework of a non-relativistic harmonic oscillator quark model. Particular attention is paid to the properties of heavy cascade baryons, $\Xi_Q$ and $\Xi'_Q$.  
\end{abstract}

\section{Introduction}

The study of heavy baryons is important to gain a better understanding of hadron structure and the strong interaction. In recent years there has been a large amount of new experimental information on heavy baryons, especially on singly heavy baryons with one charm $c$ or one bottom $b$ quark, see {\it e.g.} the PDG compilation \cite{Workman:2022ynf}. 

Recent reviews show that for heavy cascade baryons, $\Xi_{Q}$ and $\Xi'_{Q}$, all ground state $S$-wave baryons have been observed in both the charm and the bottom sector, as well as several candidates for excited $P$-wave baryons \cite{Workman:2022ynf,Chen_2023,PhysRevLett.124.222001,MoonPRD103,PhysRevD.102.071103}. 

In the bottom sector, the LHCb collaboration observed the $\Xi^-_b(6227)$ baryon in both the $\Lambda_b K$ and $\Xi_b\pi$ channels \cite{PhysRevLett.121.072002}. The $\Xi_b(6100)$ baryon which was observed by the CMS Collaboration in the $\Xi_b^-\pi^+\pi^-$ invariant mass spectrum \cite{PhysRevLett.126.252003} was recently confirmed by the LHCb Collaboration \cite{lhcbcollaboration2023observation} together with the discovery of $\Xi_b(6087)^0$ and $\Xi_b(6095)^0$. 

In this contribution we present a study of mass spectra and electromagnetic couplings of $S$- and $P$-wave $\Xi_Q$ and $\Xi'_Q$ baryons with $Q=c$ or $b$ in the framework of a non-relativistic harmonic oscillator quark model \cite{PhysRevD.105.074029,PhysRevD.107.034031,garciatecocoatzi2023decay,EOP2023,Cocoyoc2023}. 

\section{Mass spectra}

Singly heavy $qqQ$ baryons belong to either a flavor sextet ($\Sigma_Q$, $\Xi'_Q$, $\Omega_Q$) or an antitriplet ($\Lambda_Q$, $\Xi_Q$), whereas doubly heavy $QQq$ baryons constitute a flavor triplet ($\Xi_{QQ}$, $\Omega_{QQ}$) and triply heavy $\Omega_{QQQ}$ baryons a singlet. In this study we adopt a harmonic oscillator quark model to describe the spectroscopy of heavy baryons. The mass spectrum of heavy baryons is analyzed with the mass formula \cite{PhysRevD.105.074029,EOP2023}
\begin{eqnarray}
M &=& \sum_{i=1}^3 m_i +\omega_\rho n_\rho+\omega_\lambda n_\lambda+A\,S(S+1)
+B\,\frac{1}{2}\left[J(J+1)-L(L+1)-S(S+1)\right]
\nonumber\\
&& \hspace{2cm} +E\,I(I+1)+G\,\frac{1}{3}\left[p(p+3)+q(q+3)+pq\right]. 
\label{mass_formula}
\end{eqnarray}
Here $n_\rho$ and $n_\lambda$ denote the number of quanta in the $\rho$- and $\lambda$-oscillator, respectively. The ground state has $(n_\rho,n_\lambda)=(0,0)$, whereas the states with one quantum of excitation in the $\rho$ mode or the $\lambda$ mode are characterized by $(1,0)$ and $(0,1)$, respectively. The corresponding frequencies have the same spring constant $C$ but different reduced masses
\begin{eqnarray}
\omega_\rho = \sqrt{\frac{3C}{m_\rho}} = \sqrt{\frac{6C}{m_1+m_2}}, \qquad  
\omega_\lambda = \sqrt{\frac{3C}{m_\lambda}} = \omega_\rho \sqrt{\frac{m_1+m_2+m_3}{3m_3}}.
\end{eqnarray}
The labels ${L}$, ${S}$, ${J}$ and ${I}$ denote the orbital angular momentum, spin, total angular momentum and isospin, respectively. The labels $(p,q)$ represent the $SU(3)$ flavor multiplets: the flavor sextet is labeled by $(2,0) \equiv \mathbf{6}$, the anti-triplet by $(0,1) \equiv \mathbf{\bar{3}}$, the triplet by $(1,0) \equiv \mathbf{3}$, and the singlet by $(0,0) \equiv \mathbf{1}$. The parameters were determined in a fit to the experimental masses of 41 heavy baryon resonances (25 single charm, 15 single bottom and 1 double charm). A good overall fit was obtained with an root-mean square deviation of $\delta_{\rm rms}=19$ MeV \cite{EOP2023}. 

Over the last couple of years there has been a lot of progress in the understanding of single heavy cascade baryons, $\Xi_Q$ and $\Xi'_Q$. The LHCb Collaboration reported the discovery of three new resonances in the $\Lambda_c^+ K^-$ channel, $\Xi_c(2923)^0$, $\Xi_c(2939)^0$ and $\Xi_c(2965)^0$ \cite{PhysRevLett.124.222001}. The Belle Collaboration determined the spin and parity of the charmed-strange baryon, $\Xi_c(2970)^+$, to be $J^P=1/2^+$ \cite{MoonPRD103} and studied the electromagnetic decay of the excited charm baryons $\Xi_c(2790)$ and $\Xi_c(2815)$ \cite{PhysRevD.102.071103}. More recently, the LHCb Collaboration reported the discovery of 
the $\Xi_b(6087)^0$ and $\Xi_b(6095)^0$ baryons \cite{lhcbcollaboration2023observation}.

The results for single heavy cascade baryons are shown in Tables~\ref{Xic} and \ref{Xib} which are updated versions of Table~4 of Ref.~\cite{EOP2023}. All ground-state $S$-wave $\Xi'_Q$ and $\Xi_Q$ baryons have been identified. In addition, there are in total 7 $P$-wave states both for the flavor sextet and the flavor anti-triplet. We only consider states with $n_\rho+n_\lambda=0$ or 1 oscillator quanta. For ground state baryons our results are comparable to those obtained in quark model descriptions by Yoshida {\it et al.} \cite{PhysRevD.92.114029}, Karliner and Rosner \cite{PhysRevD.90.094007} and lattice QCD calculations \cite{PhysRevD.90.094507}. 

In the charm sector 5 candidates have been assigned tentatively for the flavor sextet, only for the $^2\lambda(\Xi'_c)_{1/2}$ and $^4\lambda(\Xi'_c)_{5/2}$ states there are no obvious candidates. For the anti-triplet, the $\Xi_c(2790)$ and $\Xi_c(2815)$ baryons are assigned as  $^2\lambda(\Xi_c)_J$ with $J^P=1/2^-$ and $3/2^-$, respectively. The latter assignment is based both on masses and electromagnetic decay widths \cite{PhysRevD.102.071103,PhysRevD.105.074029,EOP2023}. The $\Xi_c(2970)^+$ and $\Xi_c(3123)^+$ baryons are missing from Table~\ref{Xic}. The spin and parity of $\Xi_c(2970)^+$ were determined to be $J^P=1/2^+$ \cite{MoonPRD103} which suggests an interpretation in the harmonic oscillator quark model as an excitation of two oscillator quanta. The status of the $\Xi_c(3123)^+$ baryon is still very uncertain (one star), and in the harmonic oscillator quark model with $n_\rho+n_\lambda \leq 1$ there are no states with a comparable mass. 

In the bottom sector, there is less experimental information available \cite{Workman:2022ynf}. The $\Xi_b(6100)$ baryon which was observed by the CMS Collaboration in the $\Xi_b^-\pi^+\pi^-$ invariant mass spectrum \cite{PhysRevLett.126.252003} was recently confirmed by the LHCb Collaboration \cite{lhcbcollaboration2023observation} together with the discovery of $\Xi_b(6087)^0$ and $\Xi_b(6095)^0$. The $\Xi_b(6100)^-$ state was interpreted as a member of the isospin doublet $^2\lambda(\Xi_b)_{J=3/2}$ of the flavor anti-triplet \cite{EOP2023}. The recently discovered $\Xi_b(6095)^0$ baryon is assigned as its isospin partner, whereas the $\Xi_b(6087)^0$ baryon is assigned as a member of the isospin doublet $^2\lambda(\Xi_b)_{J=1/2}$. The corresponding theoretical masses are in good agreement with the observed values. In comparison with Table~4 of Ref.~\cite{EOP2023} here we have assigned tentatively the $\Xi_b(6327)^0$ and $\Xi_b(6333)^0$ baryons as $^2\rho(\Xi'_b)_{J}$ with $J^P=1/2^-$ and $3/2^-$, respectively. 

\begin{table}[t]
\centering
\caption{Mass spectra of single charm $S$- and $P$-wave sextet $\Xi'_c$ (top) and antitriplet $\Xi_c$ (bottom) baryons. The experimental values are taken from \cite{Workman:2022ynf,PhysRevLett.124.222001}.}
\label{Xic}
\begin{tabular}{ccccccccccc}
\hline
\noalign{\smallskip}
State & $(n_\rho,n_\lambda)$ & $I$ & $L$ & $S$ & $J^{P}$ 
& $M^{th}$(MeV) & $M^{exp}$(MeV) & Name & $\ast$ \\
\noalign{\smallskip}
\hline
\noalign{\smallskip}
$^2(\Xi'_{c})_{{1/2}}$ & (0,0) & $\frac{1}{2}$ & 0 & $\frac{1}{2}$ & $\frac{1}{2}^+$ 
& $2586 \pm 2$ & $2578.2 \pm 0.5$ & $\Xi^{\prime +}_{c}$ & 3 \\
&&&&&&& $2578.7 \pm 0.5$ & $\Xi^{\prime 0}_{c}$ & 3 \\
$^4(\Xi'_{c})_{{3/2}}$ & (0,0) & $\frac{1}{2}$ & 0 & $\frac{3}{2}$ & $\frac{3}{2}^+$ 
& $2653 \pm 2$ & $2645.1 \pm 0.3$ & $\Xi_{c}(2645)^+$ & 3 \\
&&&&&&& $2646.2 \pm 0.3$ & $\Xi_{c}(2645)^0$ & 3 \\
$^2\rho(\Xi'_{c})_{{1/2}}$ & (1,0) & $\frac{1}{2}$ & 1 & $\frac{1}{2}$ & $\frac{1}{2}^-$ 
& $3029 \pm 2$ & $3055.9 \pm 0.4$ & $\Xi_{c}(3055)^+$ & 3 \\
$^2\rho(\Xi'_{c})_{{3/2}}$ & (1,0) & $\frac{1}{2}$ & 1 & $\frac{1}{2}$ & $\frac{3}{2}^-$ 
& $3062 \pm 2$ & $3077.2 \pm 0.4$ & $\Xi_{c}(3080)^+$ & 3 \\
&&&&&&& $3079.9 \pm 1.4$ & $\Xi_{c}(3080)^0$ & 3 \\
$^2\lambda(\Xi'_{c})_{{1/2}}$ & (0,1) & $\frac{1}{2}$ & 1 & $\frac{1}{2}$ & $\frac{1}{2}^-$ 
& $2890 \pm 2$ & $\cdots$ & $\cdots$ & \\
$^2\lambda(\Xi'_{c})_{{3/2}}$ & (0,1) & $\frac{1}{2}$ & 1 & $\frac{1}{2}$ & $\frac{3}{2}^-$ 
& $2922 \pm 2$ & $2942.3 \pm 4.6$ & $\Xi_{c}(2930)^+$ & 2 \\
&&&&&&& $2938.6 \pm 0.3$ & $\Xi_{c}(2930)^0$ & 2 \\
$^4\lambda(\Xi'_{c})_{{1/2}}$ & (0,1) & $\frac{1}{2}$ & 1 & $\frac{3}{2}$ & $\frac{1}{2}^-$ 
& $2923 \pm 2$ & $2923.0 \pm 0.3$ & $\Xi_{c}(2923)^0$ & 2 \\
$^4\lambda(\Xi'_{c})_{{3/2}}$ & (0,1) & $\frac{1}{2}$ & 1 & $\frac{3}{2}$ & $\frac{3}{2}^-$ 
& $2956 \pm 2$ & $2964.9 \pm 0.3$ & $\Xi_{c}(2965)^0$ & \\
$^4\lambda(\Xi'_{c})_{{5/2}}$ & (0,1) & $\frac{1}{2}$ & 1 & $\frac{3}{2}$ & $\frac{5}{2}^-$ 
& $3011 \pm 2$ & $\cdots$ & $\cdots$ & \\
\noalign{\smallskip}
\hline
\noalign{\smallskip}
$^2(\Xi_{c})_{{1/2}}$ & (0,0) & $\frac{1}{2}$ & 0 & $\frac{1}{2}$ & $\frac{1}{2}^+$ 
& $2474 \pm 2$ & $2467.7 \pm 0.2$ & $\Xi_{c}^+$ & 3 \\
&&&&&&& $2470.4 \pm 0.3$ & $\Xi_{c}^0$ & 4 \\
$^2\rho(\Xi_{c})_{{1/2}}$ & (1,0) & $\frac{1}{2}$ & 1 & $\frac{1}{2}$ & $\frac{1}{2}^-$ 
& $2917 \pm 2$ & $\cdots$ &$\cdots$ & \\
$^2\rho(\Xi_{c})_{{3/2}}$ & (1,0) & $\frac{1}{2}$ & 1 & $\frac{1}{2}$ & $\frac{3}{2}^-$ 
& $2950 \pm 2$ & $\cdots$ & $\cdots$ & \\
$^4\rho(\Xi_{c})_{{1/2}}$ & (1,0) & $\frac{1}{2}$ & 1 & $\frac{3}{2}$ & $\frac{1}{2}^-$ 
& $2950 \pm 2$ & $\cdots$ & $\cdots$ & \\
$^4\rho(\Xi_{c})_{{3/2}}$&(1,0)&$\frac{1}{2}$ & 1 & $\frac{3}{2}$ & $\frac{3}{2}^-$ 
& $2983 \pm 2$ & $\cdots$ & $\cdots$ & \\
$^4\rho(\Xi_{c})_{{5/2}}$&(1,0)&$\frac{1}{2}$ & 1 & $\frac{3}{2}$ & $\frac{5}{2}^-$ 
& $3038 \pm 2$ & $\cdots$ & $\cdots$ & \\
$^2\lambda(\Xi_{c})_{{1/2}}$ & (0,1) & $\frac{1}{2}$ & 1 & $\frac{1}{2}$ & $\frac{1}{2}^-$ 
& $2777 \pm 2$ & $2791.9 \pm 0.5$ & $\Xi_{c}(2790)^+$ & 3 \\
&&&&&&& $2793.9 \pm 0.5$ & $\Xi_{c}(2790)^0$ & 3 \\
$^2\lambda(\Xi_{c})_{{3/2}}$ & (0,1) & $\frac{1}{2}$ & 1 & $\frac{1}{2}$ & $\frac{3}{2}^-$ 
& $2810 \pm 2$ & $2816.5 \pm 0.3$ & $\Xi_{c}(2815)^+$ & 3 \\
&&&&&&& $2819.8 \pm 0.3$ & $\Xi_{c}(2815)^0$ & 3 \\
\noalign{\smallskip}
\hline
\end{tabular}
\end{table}

\begin{table}[t]
\centering
\caption{As Table~\ref{Xic}, but for single bottom $\Xi'_b$ and $\Xi_b$ baryons. 
The experimental values are taken from \cite{Workman:2022ynf,lhcbcollaboration2023observation}.}
\label{Xib}
\begin{tabular}{ccccccccccc}
\hline
\noalign{\smallskip}
State & $(n_\rho,n_\lambda)$ & $I$ & $L$ & $S$ & $J^{P}$ 
& $M^{th}$(MeV) & $M^{exp}$(MeV) & Name & $\ast$ \\
\noalign{\smallskip}
\hline
\noalign{\smallskip}
$^2(\Xi'_{b})_{{1/2}}$ & (0,0) & $\frac{1}{2}$ & 0 & $\frac{1}{2}$ & $\frac{1}{2}^+$ 
& $5935 \pm 2$ & $5935.0 \pm 0.1$ & $\Xi'_{b}(5935)^-$ & 3 \\
$^4(\Xi'_{b})_{{3/2}}$ & (0,0) & $\frac{1}{2}$ & 0 & $\frac{3}{2}$ & $\frac{3}{2}^+$ 
& $5957 \pm 2$ & $5952.3 \pm 0.6$ & $\Xi_{b}(5945)^0$ & 3 \\
&&&&&&& $5955.3 \pm 0.1$ & $\Xi_{b}(5955)^-$ & 3 \\
$^2\rho(\Xi'_{b})_{{1/2}}$ & (1,0) & $\frac{1}{2}$ & 1 & $\frac{1}{2}$ & $\frac{1}{2}^-$ 
& $6366 \pm 3$ & $6327.3 \pm 0.4$ & $\Xi_b(6327)^0$ & 3 \\
$^2\rho(\Xi'_{b})_{{3/2}}$ & (1,0) & $\frac{1}{2}$ & 1 & $\frac{1}{2}$ & $\frac{3}{2}^-$ 
& $6373 \pm 3$ & $6332.7 \pm 0.3$ & $\Xi_b(6333)^0$ & 3 \\
$^2\lambda(\Xi'_{b})_{{1/2}}$ & (0,1) & $\frac{1}{2}$ & 1 & $\frac{1}{2}$ & $\frac{1}{2}^-$ 
& $6201 \pm 3$ & $\cdots$ & $\cdots$ & \\
$^2\lambda(\Xi'_{b})_{{3/2}}$ & (0,1) & $\frac{1}{2}$ & 1 & $\frac{1}{2}$ & $\frac{3}{2}^-$ 
& $6207 \pm 3$ & $\cdots$ & $\cdots$ & \\
$^4\lambda(\Xi'_{b})_{{1/2}}$ & (0,1) & $\frac{1}{2}$ & 1 & $\frac{3}{2}$ & $\frac{1}{2}^-$ 
& $6216 \pm 3$ & $\cdots$ & $\cdots$ & \\
$^4\lambda(\Xi'_{b})_{{3/2}}$ & (0,1) & $\frac{1}{2}$ & 1 & $\frac{3}{2}$ & $\frac{3}{2}^-$ 
& $6223 \pm 3$ & $\cdots$ & $\cdots$ & \\
$^4\lambda(\Xi'_{b})_{{5/2}}$ & (0,1) & $\frac{1}{2}$ & 1 & $\frac{3}{2}$ & $\frac{5}{2}^-$ 
& $6234 \pm 3$ & $6226.8 \pm 1.6$ & $\Xi_{b}(6227)^0$ & 3 \\
&&&&&& & $6227.9 \pm 0.9$ & $\Xi_{b}(6227)^-$ & 3 \\
\noalign{\smallskip}
\hline
\noalign{\smallskip}
$^2(\Xi_{b})_{{1/2}}$ & (0,0) & $\frac{1}{2}$ & 0 & $\frac{1}{2}$ & $\frac{1}{2}^+$ 
& $5812 \pm 2$ & $5791.9 \pm 0.5$ & $\Xi_{b}^0$ & 3 \\
&&&&&&& $5797.0 \pm 0.6$ & $\Xi_{b}^-$ & 3 \\
$^2\rho(\Xi_{b})_{{1/2}}$&(1,0)&$\frac{1}{2}$&1&$\frac{1}{2}$&$\frac{1}{2}^-$ 
& $6243 \pm 3$ & $\cdots$ & $\cdots$ & \\
$^2\rho(\Xi_{b})_{{3/2}}$&(1,0)&$\frac{1}{2}$&1&$\frac{1}{2}$&$\frac{3}{2}^-$ 
& $6249 \pm 3$ & $\cdots$ & $\cdots$ & \\
$^4\rho(\Xi_{b})_{{1/2}}$&(1,0)&$\frac{1}{2}$&1&$\frac{3}{2}$&$\frac{1}{2}^-$ 
& $6258 \pm 3$ & $\cdots$ & $\cdots$ & \\
$^4\rho(\Xi_{b})_{{3/2}}$&(1,0)&$\frac{1}{2}$&1&$\frac{3}{2}$&$\frac{3}{2}^-$ 
& $6265 \pm 3$ & $\cdots$ & $\cdots$ & \\
$^4\rho(\Xi_{b})_{{5/2}}$&(1,0)&$\frac{1}{2}$&1&$\frac{3}{2}$&$\frac{5}{2}^-$ 
& $6276 \pm 3$ & $\cdots$ & $\cdots$ & \\
$^2\lambda(\Xi_{b})_{{1/2}}$ & (0,1) & $\frac{1}{2}$ & 1 & $\frac{1}{2}$ & $\frac{1}{2}^-$ 
& $6077 \pm 3$ & $6087.2 \pm 0.5$ & $\Xi_{b}(6087)^0$  & \\
$^2\lambda(\Xi_{b})_{{3/2}}$ & (0,1) & $\frac{1}{2}$ & 1 & $\frac{1}{2}$ & $\frac{3}{2}^-$ 
& $6084 \pm 3$ & $6095.4 \pm 0.5$ & $\Xi_{b}(6095)^0$ & \\
&&&&&&& $6100.3 \pm 0.6$ & $\Xi_{b}(6100)^-$ & 3 \\
\noalign{\smallskip}
\hline
\end{tabular}
\end{table}

\section{Electromagnetic couplings}
\label{sec:em}

The assignment of quantum numbers is mostly based on energy systematics. The interpretation of the $\Xi_c(2790)$ and $\Xi_c(2815)$ resonances as $^2\lambda(\Xi_c)_J$ states with $J^P=1/2^-$ and $3/2^-$, respectively, is confirmed by an analysis of the electromagnetic decay widths \cite{EOP2023,Wang2017}. The Belle II Collaboration reported the first (and so far the only) measurement of radiative decay widths of charmed baryons. It was found that the widths of the neutral $\Xi_c(2790)^0$ and $\Xi_c(2815)^0$ baryons are large (albeit with a large uncertainty), whereas for the widths of the charged $\Xi_c(2790)^+$ and $\Xi_c(2815)^+$ baryons only an upper limit was established \cite{PhysRevD.102.071103}. 
In Table~\ref{p1} we show a comparison with other theoretical calculations. We find a reasonable agreement between our calculation and the experimental data, as well as with the $\chi$QM results \cite{Wang2017}. The decay widths of $\Xi_c(2790)$ in Ref.~\cite{Gamermann2011} in which charmed baryons are interpreted as meson-baryon molecular states are calculated to be $\sim$ 250 keV for the charged baryon and $\sim$ 120 keV for the neutral baryon. Even though these results are in qualitative agreement with the data (which have large error bars) their behavior is very different from the present results and the $\chi$QM. In the light-cone QCD sum rule approach of \cite{Aliev2019} the radiative decay width of the charged $\Xi^+_c(2790)$ baryon is calculated to be much larger than that of the neutral $\Xi^0_c(2790)$ baryon, in contradiction with the experimental data. Finally, a calculation in the relativistic quark model shows a larger radiative width for the neutral $\Xi_{c}(2815)^0$ than for the charge $\Xi_{c}(2815)^+$ state \cite{Ivanov1999}, although the calculated value of $\Gamma(\Xi_{c}(2815)^+\rightarrow\Xi_c^+ +\gamma)$ is much larger than the upper limit from experiment.

\begin{table}[t]
\caption{Radiative decay widths in keV of the $\Xi_c(2790)$ and $\Xi_c(2815)$ baryons.}
\label{p1}
\centering
\begin{tabular}{lrrccccc}
\hline
\noalign{\smallskip}
& Present & $\chi$QM \cite{Wang2017} & MB \cite{Gamermann2011} & LCQSR \cite{Aliev2019} & RQM \cite{Ivanov1999} & Exp \cite{PhysRevD.102.071103} \\
\noalign{\smallskip}
\hline 
\noalign{\smallskip}
$\Xi_{c}(2790)^{+}$ &   7.4 &   4.6 
& $249.6 \pm 41.9$ & $265 \pm 106$ && $<350$ \\
$\Xi_{c}(2790)^{0}$ & 202.5 & 263.0 
& $119.3 \pm 21.7$ & $2.7 \pm 0.8$ && $800\pm320$ \\
$\Xi_{c}(2815)^{+}$ &   4.8 &   2.8 
&&& $190 \pm 5$ & $<80$\\
$\Xi_{c}(2815)^{0}$ & 292.6 & 292.0 
&&& $497 \pm 14$ & $320 \pm 45^{+45}_{-80}$ \\
\noalign{\smallskip}
\hline
\end{tabular} 
\end{table}

\section{Summary and conclusions}
\label{sec:summary}

In summary, we presented a study of masses and electromagnetic couplings of heavy baryons in a non-relativistic harmonic oscillator quark model in which we use a G\"ursey-Radicati form for the spin-flavor dependence. The parameters are determined in a simultaneous fit to 41 heavy baryon masses, 25 single charm, 15 single bottom and 1 double charm, with an r.m.s. deviation of 19 MeV. 

The calculated masses of the single heavy cascade baryons are in good agreement with the available experimental data. A study of the radiative decay widths shows that the observed widths for the charged and neutral $\Xi_c(2790)$ and $\Xi_c(2815)$ baryons are in agreement with the assignment of these resonances as $^2\lambda(\Xi_c)_J$ states with $J^P=1/2^-$ and $3/2^-$, respectively. 

\begin{acknowledgments}
This work was supported in part by PAPIIT-DGAPA (Mexico) grant IG101423, and by CONACYT (Mexico) grant 251817. EOP acknowledges the projects (P1-0035) and (J1-3034) were financially supported by the Slovenian Research Agency. 
\end{acknowledgments}

\end{document}